\documentclass[
letterpaper,
reprint,           
twocolumn,
superscriptaddress,
amsmath,           
amssymb,           
aps,               
prl,               
notitlepage,       
longbibliography,  
floatfix,          
nofootinbib,
]{revtex4-1}
\usepackage{tabularx}
\usepackage{amsmath}
\usepackage{lipsum}
\allowdisplaybreaks[4]
\usepackage{cancel}
\usepackage{extarrows}
\usepackage{tensor}     
\usepackage{float}
\usepackage[caption = false]{subfig} 
\usepackage[final]{graphicx}   
\usepackage[
colorlinks=true,        
citecolor=blue,         
linkcolor=blue,         
urlcolor=blue           
]{hyperref}             
\usepackage{bm}         
\usepackage{xcolor}     
\usepackage{lipsum}
\usepackage{color}      
\usepackage[section]{placeins} 
\usepackage{appendix}
\usepackage{units}
\usepackage[capitalise]{cleveref}
\usepackage{units}
\newcommand{\nc}{\newcommand*}

\nc{\xbar}{\bar{x}}
\nc{\rhoeq}{\rho_{\mathrm{eq}}}
\nc{\zeq}{z_{\mathrm{eq}}}
\nc{\tla}{\tilde{\lambda}}
\nc{\bt}{\beta}
\nc{\dt}{\delta}
\nc{\Dt}{\Delta}
\nc{\vj}{\vec{j}}
\nc{\vl}{\vec{l}}
\nc{\hx}{\hat{x}}
\nc{\hy}{\hat{y}}
\nc{\bj}{\bm{j}}
\nc{\mJ}{\mathcal{J}}
\nc{\mP}{\mathcal{P}}
\nc{\Msun}{M_\odot}
\nc{\app}{\approx}
\nc{\av}[1]{\langle #1 \rangle}
\nc{\eq}[1]{Eq.~\eqref{#1}}
\nc{\al}{\alpha}
\nc{\Xstar}{X_{\ast}}
\nc{\fpbh}{f_{\mathrm{pbh}}}
\nc{\vth}{\vec{\theta}}
\nc{\vla}{\vec{\lambda}}
\nc{\vd}{\vec{d}}
\nc{\Mmin}{M_{\mathrm{min}}}
\nc{\rmd}{\mathrm{d}}
\nc{\mmin}{{m_{\mathrm{min}}}}
\nc{\mmax}{{m_{\mathrm{max}}}}
\nc{\mR}{\mathcal{R}}
\nc{\tmR}{\tilde{\mathcal{R}}}
\nc{\s}{\sigma}
\nc{\ogw}{\Omega_{\mathrm{GW}}}
\nc{\addref}{[\textcolor{red}{add ref}] }
\nc{\Om}{\Omega}
\nc{\gm}{\gamma}
\nc{\Gm}{\Gamma}
\nc{\gpcyr}{\mathrm{Gpc}^{-3}\,\mathrm{yr}^{-1}}
\nc{\Eq}[1]{Eq.~\eqref{#1}}
\nc{\Fig}[1]{Fig.~\ref{#1}}
\nc{\Table}[1]{Table~\ref{#1}}
\nc{\lvc}{LIGO/Virgo} 
\nc{\Sec}[1]{Sec.~\ref{#1}}
\nc{\eg}{\textit{e.g.~}}
\nc{\SNR}{\mathrm{SNR}}
\nc{\be}{\mathbf{\epsilon}}
\nc{\bn}{\mathbf{n}}
\nc{\bd}{\mathbf{d}}
\nc{\ba}{\mathbf{a}}
\nc{\eps}{\epsilon}
\nc{\bnu}{\mathbf{\nu}}
\nc{\mb}{\mathbf}
\nc{\bbt}{\mathbf{t}}
\nc{\bth}{\mathbf{\theta}}
\nc{\bep}{\mathbf{\epsilon}}
\nc{\uni}{\mathrm{U}}
\nc{\logu}{\operatorname{\mathrm{log-U}}}
\nc{\RN}{\mathrm{RN}}
\nc{\BN}{\mathrm{BN}}
\nc{\GN}{\mathrm{GN}}
\nc{\mcN}{\mathcal{N}}
\nc{\GWB}{\mathrm{GW}}
\nc{\yr}{\mathrm{yr}}
\nc{\Am}{\mathcal{A}}
\nc{\Dm}{\mathcal{D}}
\nc{\Hm}{\mathcal{H}}
\nc{\sovast}{Soviet Ast.}

\nc{\mrm}{\mathrm}
\nc{\BE}{B\scriptsize{AYES}\normalsize{E}\scriptsize{PHEM}\normalsize  }

\nc{\Ostgw}{\Omega_{\mathrm{GW}}^{\mathrm{ST}}}
\nc{\Ottgw}{\Omega_{\mathrm{GW}}^{\mathrm{TT}}}
\nc{\Ovlgw}{\Omega_{\mathrm{GW}}^{\mathrm{VL}}}
\nc{\Oslgw}{\Omega_{\mathrm{GW}}^{\mathrm{SL}}}
\nc{\cosxi}{\beta}

\nc{\gmPL}{\gamma_{\mathrm{PL}}}
\nc{\APL}{A_{\mathrm{PL}}}

\def\({\left(}
\def\){\right)}
\def\[{\left[}
\def\]{\right]}

\def\e{\begin{equation}}
\def\q{\end{equation}}
\def\m{\begin{eqnarray}}
\def\n{\end{eqnarray}}
\nc{\red}[1]{\textcolor{red}{#1}}

\begin{document}


\title{Cosmological Interpretation for the Stochastic Signal in Pulsar Timing Arrays}
\author{Yu-Mei Wu}
\email{ymwu@ucas.ac.cn} 
\affiliation{School of Fundamental Physics and Mathematical Sciences, Hangzhou Institute for Advanced Study, UCAS, Hangzhou 310024, China}
\affiliation{School of Physical Sciences, 
    University of Chinese Academy of Sciences, 
    No. 19A Yuquan Road, Beijing 100049, China}

\author{Zu-Cheng Chen}
\email{Corresponding author: zucheng.chen@bnu.edu.cn}
\affiliation{Department of Astronomy, Beijing Normal University, Beijing 100875, China}
\affiliation{Advanced Institute of Natural Sciences, Beijing Normal University, Zhuhai 519087, China}
\affiliation{Department of Physics and Synergistic Innovation Center for Quantum Effects and Applications, Hunan Normal University, Changsha, Hunan 410081, China}

\author{Qing-Guo Huang}
\email{Corresponding author: huangqg@itp.ac.cn}
\affiliation{School of Fundamental Physics and Mathematical Sciences, Hangzhou Institute for Advanced Study, UCAS, Hangzhou 310024, China}
\affiliation{School of Physical Sciences, 
    University of Chinese Academy of Sciences, 
    No. 19A Yuquan Road, Beijing 100049, China}
\affiliation{CAS Key Laboratory of Theoretical Physics, 
    Institute of Theoretical Physics, Chinese Academy of Sciences,Beijing 100190, China}

\begin{abstract}
The pulsar timing array (PTA) collaborations have recently reported compelling evidence for the presence of a stochastic signal consistent with a gravitational-wave background. 
In this letter, we combine the latest data sets from NANOGrav, PPTA and EPTA collaborations to explore the cosmological interpretations for the detected signal from first-order phase transitions, domain walls and cosmic strings, separately. We find that the first-order phase transitions and cosmic strings can give comparable interpretations compared to supermassive black hole binaries (SMBHBs) characterized by a power-law spectrum, but the domain wall model is strongly disfavored with the Bayes factor compared to the SMBHB model being 0.009. Furthermore, the constraints on the parameter spaces indicate that: 1) a strong phase transition at temperatures below the electroweak scale is favored and the bubble collisions make the dominant contribution to the energy density spectrum; 2) the cosmic string tension is $G \mu \in [1.46, 15.3]\times 10^{-12}$ at $90\%$ confidence interval and a small reconnection probability $p<6.68\times 10^{-2}$ is preferred at $95\%$ confidence level, implying that the strings in (super)string theory are strongly favored over the classical field strings. 

\end{abstract}
\maketitle

\textbf{Introduction.} After the detection of the gravitational waves (GWs) from compact binary coalescences by the ground-based detectors \citep{LIGOScientific:2016aoc,LIGOScientific:2018mvr,LIGOScientific:2020ibl,LIGOScientific:2021djp}, one of the most anticipated GW sources is the stochastic gravitational-wave background (SGWB) promisingly captured by the pulsar timing arrays (PTAs)~\cite{1978SvA....22...36S,Detweiler:1979wn}. Several individual PTA collaborations, including North American Observatory for Gravitational Waves  (NANOGrav) \citep{McLaughlin:2013ira}, Parkes PTA (PPTA) \citep{Manchester:2012za}, European PTA (EPTA) \citep{Kramer:2013kea}, and the joint collaboration International PTA (IPTA) \cite{Manchester:2013ndt}, have been endeavoring to search for the SGWB with increasing sensitivity, by accumulating more than a decade's timing data from dozens of pulsars. The emerging Chinese PTA (CPTA)~\cite{2016ASPC..502...19L}, Indian PTA (InPTA)~\cite{Tarafdar:2022toa}, and MeerKAT PTA~\cite{Miles:2022lkg} are also making significant contributions. Recently, NANOGrav~\cite{NANOGrav:2023hde,NANOGrav:2023gor}, PPTA~\cite{Zic:2023gta,Reardon:2023gzh}, EPTA~\cite{Antoniadis:2023lym,Antoniadis:2023ott}, and CPTA~\cite{Xu:2023wog} all have found evidence supporting the existence of a stochastic signal consistent with the Hellings-Downs~\cite{Hellings:1983fr} inter-pulsar correlations, pointing to the GW origin of the signal.

\begin{figure*}[tbp!]
	\centering
	\includegraphics[width = 1.0\textwidth]{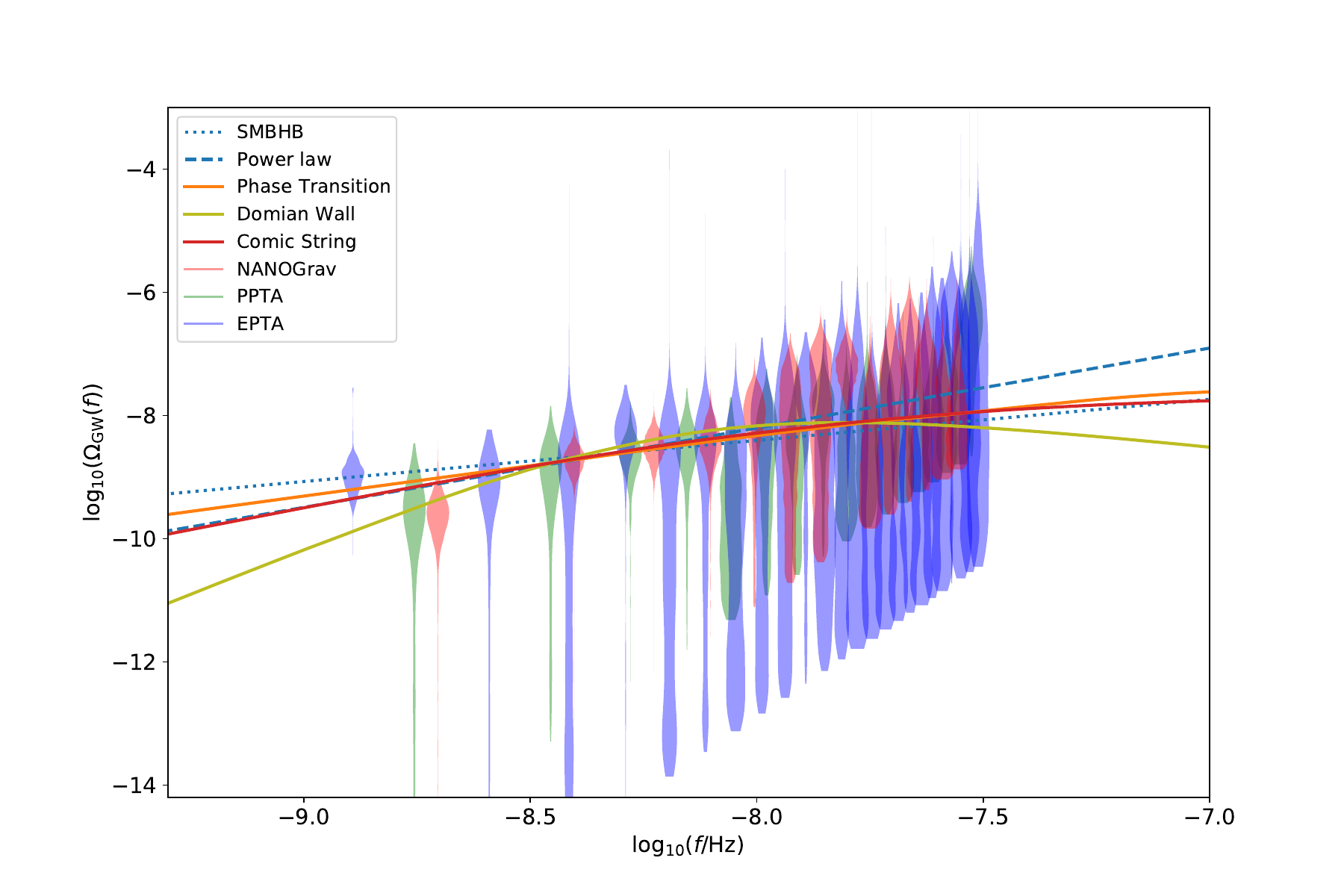}\caption{
	The SGWB spectra for different models with the median value of parameters. The magenta, green and purple violin plots show the free spectrum posteriors in the analyses from the NANOGrav 15-yr data set, PPTA DR3, and EPTA DR2.
	}
\label{fig:spectrum}
\end{figure*}

The next critical task is to identify the origin of the signal. It is believed that the large population of supermassive black hole binaries (SMBHBs) produce the brightest SGWB source at nanoHertz \citep{1995ApJ...446..543R,Jaffe:2002rt,Burke-Spolaor:2018bvk}. If all contributing SMBHBs are inspiraling in circular orbits and their orbital evolution is dominated by gravitational radiation, the timing residual power spectral density induced by the SGWB can be very well modeled by a simple power-law $f^{13/3}$ \citep{Phinney:2001di}. 
Although the power-law spectral shape used to model the observed excess low-frequency residual powers in NANOGrav 15-year data set deviates somewhat from the predicted value of $13/3$, the discrepancy can be explained by considering a more realistic scenario for SMBHBs
accounting for the effects of galactic environmental processes, such as dynamical friction and stellar scattering \citep{Kelley:2016gse}, possible significant eccentricity in the orbits of SMBHBs \citep{Chen:2016zyo}, and the intrinsic discrete nature of the sources \citep{Sesana:2008mz}.
In fact, further investigations have been conducted on SMBHB sources in the NANOGrav 15-year data set, utilizing astrophysically informed models \citep{NANOGrav:2023hfp,Ghoshal:2023fhh,Shen:2023pan,Ellis:2023dgf,Bi:2023tib}.

Although the SGWB from SMBHBs is supposed to be the most promising source for the signal detected by the PTAs, the signal at nanoHertz can also originate from some cosmological processes \citep{Maggiore:1999vm, Caprini:2018mtu,Chen:2021wdo,Wu:2021kmd,Chen:2021ncc,PPTA:2022eul,Wu:2023pbt,Wu:2023dnp,Madge:2023cak}, such as cosmological first-order phase transition \citep{Kibble:1976sj, Vilenkin:1984ib,Caprini:2010xv, Kobakhidze:2017mru, Arunasalam:2017ajm, Xue:2021gyq, NANOGrav:2021flc,Moore:2021ibq,Addazi:2023jvg,Athron:2023xlk,Bringmann:2023opz}, cosmic strings \citep{Damour:2004kw,Siemens:2006yp, Chen:2022azo,Bian:2022tju}, domain walls \citep{Ferreira:2022zzo}, and scalar-induced GW \citep{tomita1967non,Saito:2008jc,Young:2014ana,Yuan:2019udt,Yuan:2019wwo,Chen:2019xse,Cai:2019bmk,Yuan:2019fwv,Liu:2021jnw,Liu:2023ymk,Cai:2023dls} accompanying the formation of primordial black holes~\cite{Zeldovich:1967lct,Hawking:1971ei,Carr:1974nx,Chen:2018czv,Chen:2018rzo, Liu:2018ess,Liu:2019rnx,Chen:2019irf,Liu:2020cds,Wu:2020drm,Chen:2021nxo,Chen:2022fda,Chen:2022qvg,Liu:2022iuf,Zheng:2022wqo}. Each of these predicted sources exhibits a distinct spectral shape in the PTA frequency band. 
The NANOGrav and EPTA collaborations have already searched for signals from these new physics in their respective latest data sets~\citep{NANOGrav:2023hvm, Antoniadis:2023xlr}. Some recent studies have also explored the non-astrophysical interpretations of the SGWB signal and their potential implications for individual data set \citep{Han:2023olf,Li:2023bxy,Kitajima:2023cek,Murai:2023gkv,Athron:2023mer}. In this letter, we aim to further investigate cosmological scenarios using the combined data sets from NANOGrav, PPTA, and EPTA collaborations, with the goal of breaking the degeneracy among these models.
\Fig{fig:spectrum} presents a comparison between the joint PTA posteriors on the GW energy density $\Omega_{\rm{GW}}(f)$ and the spectra from both astrophysical and cosmological sources. It is noteworthy that except the domain-wall model, all other models share comparable consistency with the data. Hence, cosmological sources could serve as an alternative interpretation for the detected signal, and the implications for the fundamental physics underlying \Fig{fig:spectrum} will be explored.

\begin{figure}[!tbp]
    \centering
	\includegraphics[width=1.0\linewidth]{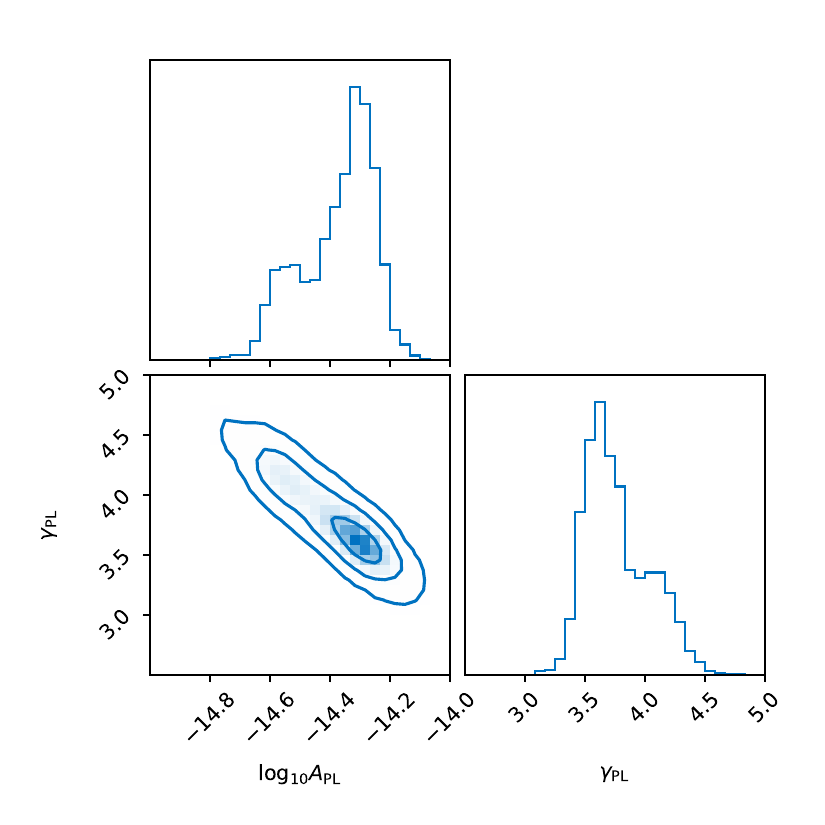}
		\caption{Corner plot showing the one- and two-dimensional posterior distributions for the power-law model. The two dimensional plots are shown in $1\sigma$, $2\sigma$ and  $3\sigma$ contours.}
		\label{PL_post}
\end{figure}

\textbf{Data analyses.}
The spectrum of an isotropic SGWB can be described by the dimensionless GW energy density parameter per logarithm frequency,
\e
\Om_{\rm{GW}}(f)=\frac{1}{\rho_c}\frac{d \rho_{\rm{GW}}}{d \ln f},
\label{Om_gw}
\q
where $\rho_c=3H_0^2 /8\pi G$ is the critical energy density of the Universe. When the Hellings-Downs correlations~\cite{Hellings:1983fr} reflect the geometric property of the quadruple nature in an SGWB, the energy spectrum will yield the information about the source of the SGWB. For example, the power-law spectrum predicted by the SMBHB sources takes the form of \citep{Thrane:2013oya}
\e
\Om_{\rm{PL}}(f) =\frac{2\pi^2\APL^2}{3H_0^2}\(\frac{f}{f_{\yr}}\)^{5-\gmPL}f_{\yr}^2,
\q
where $\APL$ is the amplitude of the GW characteristic strain measured at $f_{\yr}=1/\rm{year}$ and $\gmPL$ is the power-law index with the expected value of $\gamma=13/3$. While the NANOGrav 15-yr data set prefers a shallower slope than the excepted value, the combined data sets are more compatible with the prediction, see the posteriors of the power-law model in \Fig{PL_post}. The $5\%$ and $95\%$ quantiles for the model parameters are: $\log_{10} \APL \in [2.52,6.02]\times 10^{-15}$ and $\gmPL \in [3.42, 4.27]$. In addition, the Bayes factor (see \Eq{BF}) between the power-law model (with $\gamma$ varied) and SMBHB model (with $\gamma=13/3$) is about $0.57$, indicating that these two models are comparable. Hence, when we compare the fitness to data of different models, we still use the highly expected SMBHB sources as the fiducial model.

\vspace{10pt}

NANOGrav, PPTA and EPTA collaborations all analyzed their data sets by using a free spectrum which allows the amplitude of GW spectrum in each frequency bin to vary independently. 
Following the methodology outlined in \citep{Moore:2021ibq}, we utilize the posterior distribution of the free spectrum with Hellings-Downs correlations to conduct a Hierarchical Bayesian analysis on the SGWB parameters for various models. Initially, we construct kernel density estimates for the posterior distribution of $\Omega_{\rm{GW}}(f_{i})$ at each frequency using the latest NANOGrav, PPTA, and EPTA data sets.
Subsequently, for each candidate model, we compute the logarithmic probability density functions (log-PDFs) at different frequencies by utilizing the kernel density estimates. These log-PDFs are then summed to obtain the overall log-likelihood function.
We use the \texttt{Bilby}~\cite{Ashton:2018jfp,Romero-Shaw:2020owr} package to evaluate the likelihood and employ the \texttt{dynesty}~\cite{Speagle:2019ivv} implementation to perform the sampling needed for parameter estimations.

Bayesian model comparison is employed to assess which model is more favored by the available data. In this letter, we adopt the SMBHB model as the fiducial model, and calculate the Bayes factor for a particular cosmological source $\mathcal{M}_{\rm{XX}}$ against the SMBHB model $\mathcal{M}_{\rm{SMBHB}}$ as
\e\label{BF}
\rm{BF}_{SMBHB}^{XX}=\frac{\rm{Pr}(\mathcal{D}/\mathcal{M}_{XX})}{\rm{Pr}(\mathcal{D}/\mathcal{M}_{SMBHB})},
\q
where $\rm{Pr}(\cal{D}/\cal{M})$ is the evidence that measures the probability to obtain the data $\cal{D}$ under the hypothesis of model $\cal{M}$. According to interpretation of the Bayes factor \citep{BF}, if $0.33\le \rm{BF}_{SMBHB}^{XX}\le 3$, then the evidence supporting $\mathcal{M}_{\rm{XX}}$ over $\mathcal{M}_{\rm{SMBHB}}$ ($\rm{BF}_{SMBHB}^{XX}>1$) or 
$\mathcal{M}_{\rm{SMBHB}}$ over $\mathcal{M}_{\rm{XX}}$ ($\rm{BF}_{SMBHB}^{XX}<1$) is ``not worth more than a bare mention", while when $0.05\le \rm{BF}_{SMBHB}^{XX}\le 20$, the model with lower evidence is strongly disfavored.

In the following parts, we will discuss the  cosmological SGWB sources including the first-order phase transitions, domain walls and cosmic strings, respectively. We will report their respective Bayes factors compared to the SMBHB model and discuss their physical implications through the parameter-space exploration. For later convenience, we summarize the parameters and their priors for each model in \Table{prior}.

\begin{table}[!htbp]
    \footnotesize
    \caption{Parameters and their prior distributions used in the analyses. Here U and log-U represent a uniform and log uniform distribution, respectively.}
    \label{prior}
    \begin{tabular}{c c c }
        \hline
        Parameters & Description & Prior  \,\\
        \hline
        \multicolumn{3}{c}{\textbf{Power Law}}\,\\
        $A_{\rm{PL}}$ & power-law amplitude &$\logu[-16,-13]$  \\
        $ \gamma_{\rm{PL}}$ & power-law index & $\rm{U}[1, 7]$  \, \\
        \hline
        \multicolumn{3}{c}{\textbf{Phase Transition}}\,\\	        
        $T_n [\rm{GeV}]$ & phase transition temperature & $\logu[-4, 3]$  \\
        $\al_{\rm{PT}}$ & phase transition strength & $\logu[-2, 1]$  \\
        $H_n/\bt$ & bubble nucleation rate & $\logu[-2,0]$  \\
        $\eta$ & friction coefficient & $\logu[-2,1]$\\
        \hline
        \multicolumn{3}{c}{\textbf{Domain Wall}}\,\\
        $T_{a} [\rm{GeV}]$ & annihilation temperature &$\logu[-2.57, 10]$  \\
        $ \al_{\rm{DW}}$ & energy fraction in domain walls & $\logu[-3, -0.5]$  \, \\
        b & high frequency spectral index &$\rm{U}[0.5,1]$  \\
        c & spectrum width & $\rm{U}[0.3, 3]$  \\
        \hline
        \multicolumn{3}{c}{\textbf{Cosmic String}}\,\\
        $G \mu$ & cosmic string tension &$\logu[-15, -8]$ \, \\
        $p$ &reconnection probability &$\logu[-3,0]$  \\
        \hline
    \end{tabular}
\end{table}

\textbf{SGWB from first-order phase transitions.} Some extensions of the Standard Model predict the occurrence of a first-order phase transition \cite{Kibble:1976sj,Vilenkin:1984ib,Fromme:2006cm,Schwarz:2009ii,Schwaller:2015tja}. It happens when the temperature drops to some level, the original symmetry is broken, and the true vacuum state with the lower energy condensate as bubbles in the plasma which is still in the false vacuum background. These bubbles absorb energy from the false vacuum which turns into the kinetic energy of the bubble walls, and expand in the false vacuum. The collisions between nearby bubbles and the interaction between bubbles and the surrounding plasma will produce GWs.

There are three main sources of GWs orginated from first-order phase transitions: (i) collisions of bubble walls; (ii) collisions of sound waves in the plasma; (iii) turbulence in the plasma. Because the turbulence usually contributes subordinately compared with the sound waves, we do not include it in this work. The contribution from the sound waves to GW spectrum is \citep{Hindmarsh:2017gnf}

\e
\begin{aligned}
h^2\Om_{\rm{PT}}^{\rm{SW}}(f)=& 1.8\times 10^{-5}v_{w}\(\frac{\kappa_{\rm{sw}} \al_{\rm{PT}}}{1+\al_{\rm{PT}}}\)^2\(\frac{H_n}{\bt}\) \(\frac{10}{g_{*}}\)^{1/3} \\
&\times \(\frac{f}{f_{\rm{sw}}}\)^3\(\frac{7}{4+3(f/f_{\rm{sw}})^2}\)^{7/2} \Upsilon(\tau_{\rm{sw}}),
\end{aligned}
\q
where $v_{w}$ is the bubble wall velocity; $\al_{\rm{PT}}$ is the strength of the phase transition; $\kappa_{\rm{sw}}$ is the fraction of the vacuum energy transferred into the kinetic energy of plasma and depends on $v_w$ and $\alpha_{\rm{PT}}$ \citep{Espinosa:2010hh}; $\beta/H_n$ is the bubble nucleation rate; $g_*$ is the effective number of relativistic degrees of freedom and takes different values at different nucleation temperature $T_{n}$, i.e, $g_*\approx100$ for $T_{n}>0.2 \rm{GeV}$, $g_*\approx10$ for $0.1{\rm{MeV}}<T_{n}<0.2 {\rm{GeV}}$ and $g_*\approx 3$ for $T_{n}<0.1 \rm{MeV}$ \citep{Husdal:2016haj}; $\Upsilon(\tau_{\rm{sw}})=1-(1+2\tau_{\rm{sw}}H_n)^{-1/2}$ is a suppression factor accounting for the effect of the finite lifetime of the sound wave \citep{Guo:2020grp}, which is approximately given by $\tau_{\rm{sw}}\approx R_{n}/\bar{U}_{f}$ \citep{Weir:2017wfa}, with the average bubble separation $R_{n}=(8\pi)^{1/3}\bt^{-1}{\rm{Max}}(v_{w},c_s)$ and root-mean-square fluid velocity $\bar{U}_{f}=\sqrt{3 \kappa_{\rm{sw}}\al_{\rm{PT}}/[4(1+\al_{\rm{PT}})]}$ \citep{Hindmarsh:2015qta,Caprini:2019egz}. The value of the peak frequency $f_{\rm{sw}}$ is given by \citep{Weir:2017wfa}
\e
f_{\mathrm{sw}}\approx6.1\times 10^{-10} {\rm{Hz}} \(\frac{1}{v_{w}}\)\(\frac{\bt}{H_n}\)\(\frac{T_n}{\rm{10MeV}}\)\(\frac{g_*}{10}\)^{1/6}.
\q
Meanwhile, the contribution from the bubble collisions to GW spectrum is given by  \cite{Jinno:2016vai}
\e
\begin{aligned}
h^2\Om_{\rm{PT}}^{\rm{BC}}(f)\!\!=& 3.6\times 10^{-5}\Delta(v_{w})\(\frac{\kappa_{\phi} \al_{\rm{PT}}}{1+\al_{\rm{PT}}}\)^2\(\frac{H_n}{\bt}\)^2 \(\frac{10}{g_{*}}\)^{1/3} \\
&\times S(f/f_{\rm{bc}}),
\end{aligned}
\q
where $\Delta(v_{w})=0.48 v_w^3/(1+5.3v_w^2+5v_w^4)$ \citep{Jinno:2016vai} and $\kappa_{\phi}$ is the efficiency of the vacuum energy transformed directly into the field. 
The spectral shape $S(x)$ can be parameterized as 
\e
S(x)=\frac{(a+b)^{c}}{(b x^{-\frac{a}{c}}+a x^{\frac{b}{c})^c}}.
\label{S_x}
\q
The parameters $a$, $b$, $c$ vary with models and the derivation methods including the envelope approximation, semi-analytic approach and lattice simulation \citep{Jinno:2016vai,Lewicki:2020azd,Cutting:2020nla}. In this letter, we take $a=1$, $b=2.2$, $c=2$
allowed in the semi-analytic method \citep{Lewicki:2020azd}. 
The peak frequency locates at \citep{Jinno:2016vai},
\e
f_{\rm{bc}}\approx1.1\times 10^{-9} {\rm{Hz}} \(\frac{f_{*}}{\bt}\)\(\frac{\bt}{H_n}\)\(\frac{T_n}{\rm{10MeV}}\)\(\frac{g_*}{10}\)^{1/6}
\q
and $f_*/\bt\approx0.1$ from the  semi-analytic method \citep{Lewicki:2020azd}.

The relative contributions from the sound waves and bubble collisions to the GW spectrum depend strongly on the dynamics of the phase transition~\citep{Espinosa:2010hh}. In the ``non-runaway" scenario where bubbles expanding in the plasma can reach a terminal velocity and most of the energy released during the phase transition transfers to the surrounding plasma through its interaction with the expanding walls, the sound wave contribution dominates the GW spectrum. However, 
if the released energy is so large that the friction between the bubble walls and the plasma cannot prevent the walls from accelerating, the runaway scenario is reached, and the bubble collision could also contribute significantly to the GW spectrum.
Based on this picture, we introduce a friction efficiency parameter~\citep{Espinosa:2010hh}, $\eta$, to describe the interaction strength between the bubbles and surrounding plasma. We also relate the bubble wall velocity $v_{w}$ and the efficiency factor $\kappa_{\rm{sw}}$ and $\kappa_{\phi}$, to the parameters $\alpha_{\rm{PT}}$ and $\eta$, which will determine the contributions of plasma and bubbles to the energy budget.

\begin{figure}[!tbp]
	\centering
	\includegraphics[width=\linewidth]{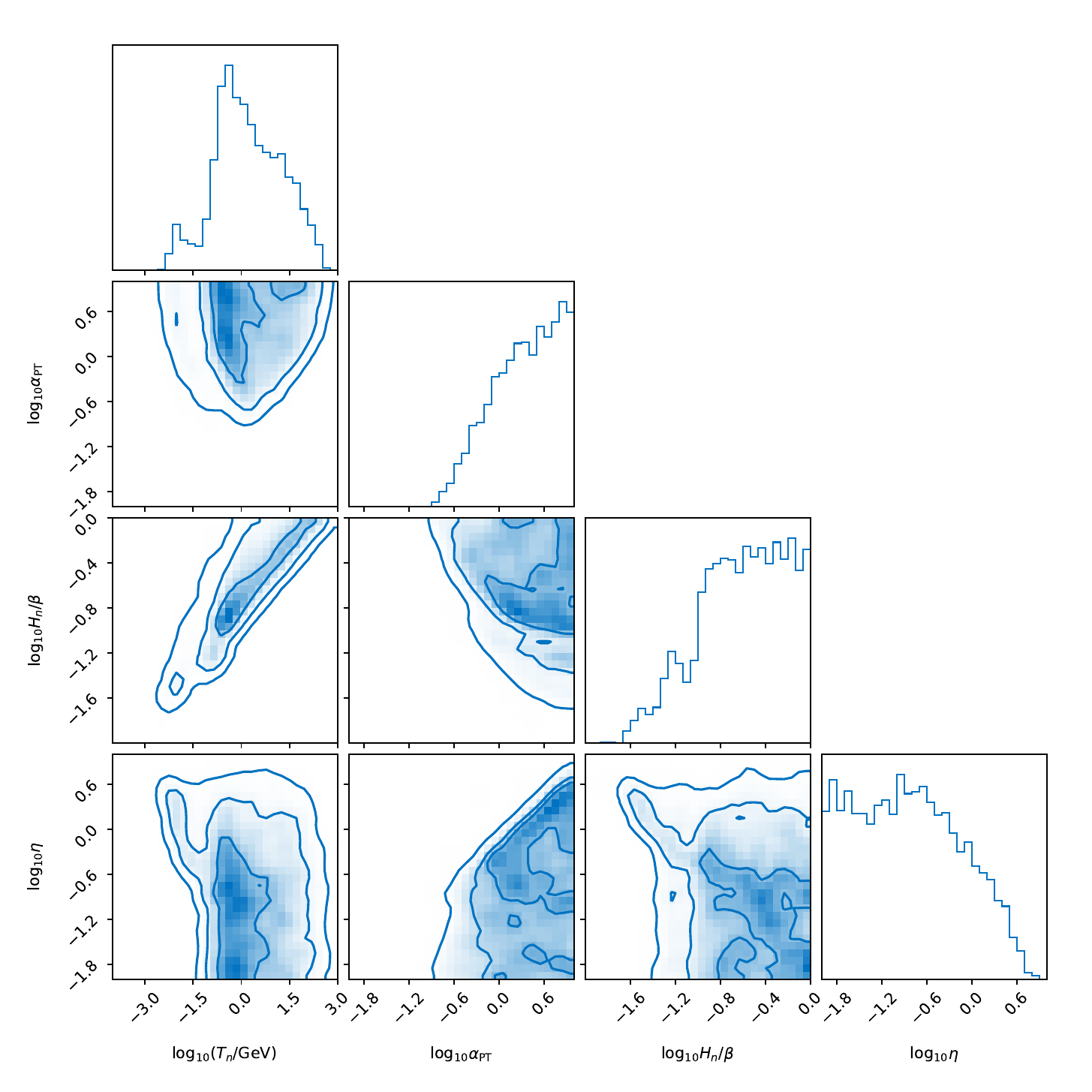}
	\caption{\label{PT_post} Corner plot showing the one- and two-dimensional posterior distributions for the model of first-order phase transitions. The two dimensional plots are shown in $1\sigma$, $2\sigma$, and $3\sigma$ contours. }
\end{figure}

The Bayes factor of the first-order phase transition model versus the SMBHB model is $\rm{BF}^{\rm{PT}}_{\rm{SMBHB}}=0.799$, indicating that this cosmological interpretation receive comparable support from the current data as the SMBHB model.
The posterior distributions for the parameters in the model of first-order phase transitions are shown in \Fig{PT_post}. The $5\%$ and $95\%$ quantiles for the model parameters are: $T_{n}\in \unit[[2.46\times 10^{-2}, 9.27]]$\,GeV, $\al_{\rm{PT}}\in [0.35,8.84]$, $H_{n}/\bt \in [0.048,0.83]$, and $\eta \in [0.013,2.17]$.
The results suggest that a strong phase transition ($\al_{\rm{PT}}>0.1$) is favored, and the bubble collisions dominate the phase transition process ($\eta<1$). 

\vspace{10pt}

\vspace{10pt}

\textbf{SGWB from domain walls.} Domain walls are lamellar topological defects formed in the early Universe when a discrete symmetry is broken \citep{Vilenkin:1982ks}. They arise in various well-motivated particle physics frameworks, such as Higgs models \cite{Battye:2020jeu}, supersymmetry \citep{Kovner:1997ca}, grand unification \citep{Lazarides:1981fv}. However, domain wall networks cannot stay stable and begin annihilating before they dominate over the total energy density of the Universe~\citep{Zeldovich:1974uw}. The motions and annihilations of the domain walls accompany with time-varying quadruples and thus act as an efficient source of GWs.

\begin{figure}[!htbp]
    \centering
	\includegraphics[width=1.0\linewidth]{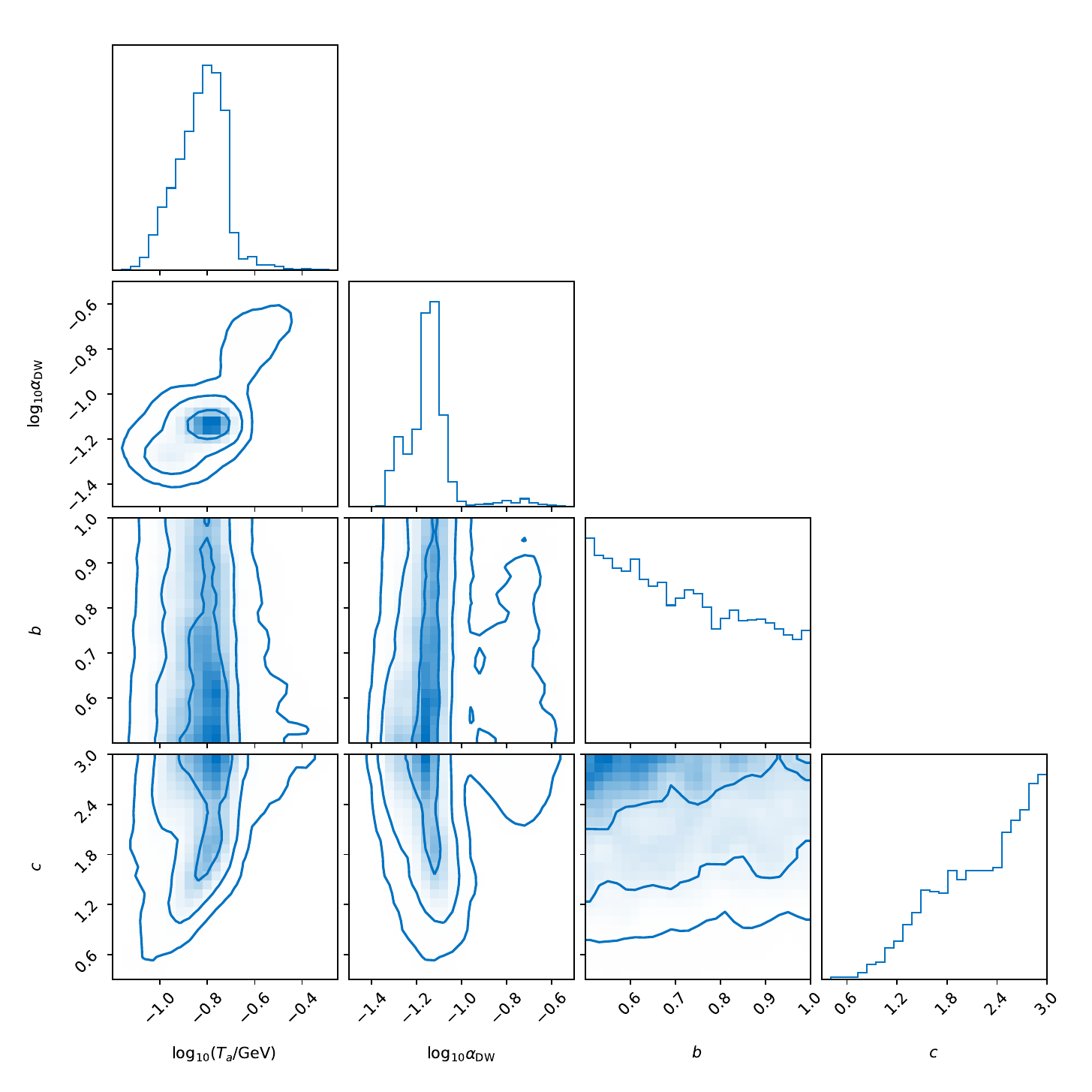}
		\caption{Corner plot showing the one- and two-dimensional posterior distributions for the model of domain walls . The two dimensional plots are shown in $1\sigma$, $2\sigma$ and  $3\sigma$ contours. }
		\label{DW_post2}
\end{figure}

The energy spectrum from the domain walls in a model-independent search can be expressed as \citep{Ferreira:2022zzo}
\e
\begin{aligned}
h^2\Om_{\rm{DW}}(f)=&  10^{-10}\tilde{\eps}\(\frac{ \al_{\rm{DW}}}{0.01}\)^2\(\frac{10}{g_{*}}\)^{1/3} S_{\rm{DW}}(f/f_{\rm{dw}}).
\end{aligned}
\q
The peak frequency locates at
\e
f_{\rm{dw}}=10^{-9}{\rm{Hz}}\(\frac{g_*}{10}\)^{1/6}\(\frac{T_{a}}{10 \rm{MeV}}\),
\q
where $\al_{\rm{DW}}=\rho_{\rm{DW}}/\rho_{\rm{tot}}$ is the fraction of total energy density in domain walls at the annihilation temperature $T_a$; $\tilde{\eps}$ is an efficient parameter to be extracted from numerical simulations and we fix $\tilde{\eps}=0.7$~\citep{Hiramatsu:2013qaa}. The spectral shape can also take the parameterized form of \Eq{S_x}. The causality requires that the slope of the spectrum approximates $\Om_{\rm{DW}}\propto f^3$  when $f<f_{\rm{dw}}$, so we fix $a=3$. Although numerical simulations suggest that $b\approx c\approx 1$, we set $b$ and $c$ to be free parameters following \citep{Ferreira:2022zzo}, 

The Bayes factor of the domain-wall model versus the SMBHB model is $\rm{BF}^{\rm{DW}}_{\rm{SMBHB}}=0.009$, suggesting that domain walls are strongly disfavored and hence are unlikely to act as an viable interpretation of the signal in the PTA data sets.
The posterior distributions for the parameters in the model of domain walls are  shown in \Fig{DW_post2}. It is clear the parameter space is highly compressed, and the $5\%$ and $95\%$ quantiles for the model parameter are: $T_{a}\in \unit[[9.98, 197]]\ {\rm{MeV}}$, $\al_{\rm{DW}}\in [0.051,0.096]$, $b \in [0.518,0.96]$, and $c \in [1.22,2.95]$.
Although the constraint $\al_{\rm{DW}}<0.3$ ensures that there are no deviations from radiation domination, and $T_{a}>\unit[2.7]{{\rm{MeV}}}$ guarantees that Big Bang Nucleosynthesis (BBN) is not affected, the spectral width $c>1.22$ has a conflict with the numerical simulation of $c\approx 1$~\citep{Hiramatsu:2013qaa}.

\vspace{10pt}

\textbf{SGWB from cosmic strings.} Comic strings are one-dimensional topological defects that may result from the spontaneous symmetry breaking happened in the phase transitions \citep{Kibble:1976sj,Vilenkin:1984ib}. 
They can be the field strings that are predicted by well-motivated inflationary models~\citep{Jeannerot:2003qv}, and can also be the fundamental strings in (super)string theory and naturally arise in a brane inflation scenario~\citep{Sarangi:2002yt,Dvali:2003zj}. 
With huge string tension, the cosmic strings are dramatically highly relativistic. When two cosmic strings collide in the three-dimensional space, they can reconnect with a characteristic probability $p$ and form loops. 
While the field strings always reconnect when they meet and exchange partners and hence take $p=1$, the strings in (super)string theory are  predicted to take a smaller reconnection probability, e.g., $10^{-3}<p<1$ for fundamental strings and $0.1<p<1$ for Dirichlet strings \citep{Jackson:2004zg}, because they are actually moving in a higher-dimensional space. Once the loops are formed, they start oscillating and shrinking in size through radiating GWs \citep{Vilenkin:1981bx}.

The energy density spectrum from the cosmic string network for both the classical field strings and superstrings can be characterized by the dimensionless string tension $G \mu$ and reconnection probability $p$ \citep{Blanco-Pillado:2017oxo}, 
\e
\Om_{\rm{CS}}=\frac{8\pi G f}{3H_0^2 p}(G\mu)^2\sum_{k=1}^{\infty} C_k P_k,
\q
where
\e
C_k(f)=\int_0^{t_0}\frac{dt}{(1+z)^5}\frac{2n}{f^2}n(l,t).
\q
Here $z$ is the redshift, $k$ labels the harmonic modes of cosmic-string loops,  $P_k$ is the radiation power spectrum of each loop, and $n(l,t)$ is the number of loops per unit volume per unit range of loop length $l$ existing at time $t$. The SGWB from a network of cosmic strings has been computed in \citep{Blanco-Pillado:2017oxo} and the output of the expected energy density spectrum has been publicly available\footnote{\url{http://cosmos.phy.tufts.edu/cosmic-string-spectra/}}.

\begin{figure}[!tbp]
	\centering
		\includegraphics[width=1.0\linewidth]{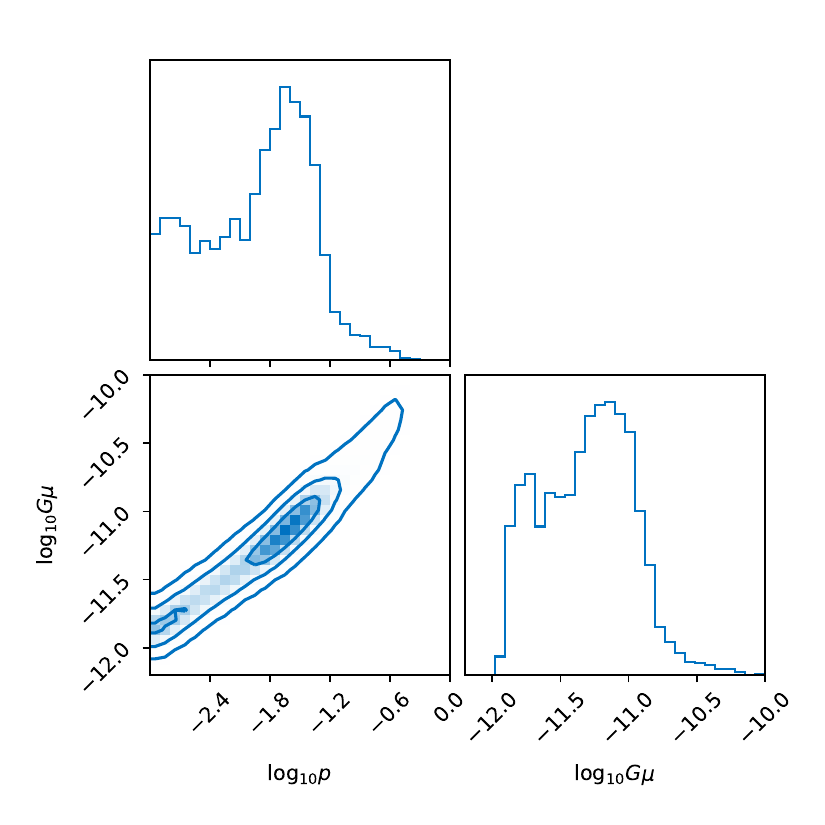}
		\caption{Corner plot showing the one- and two-dimensional posterior distributions for the model of cosmic strings. The two dimensional plots are shown in $1\sigma$, $2\sigma$ and  $3\sigma$ contours.}
		\label{CS_post}
\end{figure}

In the analyse, the prior of the reconnection probability is set as $p \in \logu[-3,0]$ to align with the constraints from the fundamental strings. The Bayes factor of the cosmic-string model versus the SMBHB model is $\rm{BF}^{\rm{CS}}_{\rm{SMBHB}}=1.699$, indicating that the cosmic strings are also a viable source of the PTA signal. The posterior distributions for the parameters are shown in \Fig{CS_post}. The $95\%$ upper limit of the reconnection probability $p$ is $6.68\times 10^{-2}$, and the $5\%$ and $95\%$ quantiles for the cosmic string tension are $G \mu \in [1.46,15.3]\times 10^{-12}$.
The result that a smaller reconnection probability is more favored indicates that if the detected signal in the PTA data sets originate from the cosmic strings, it should come from strings in (super)strings theory and is unlikely of the classical field strings.

\vspace{10pt}

\textbf{Conclusion and discussion.} 
In this letter, we combine the NANOGrav-15yr data set, PPTA DR3 and EPTA DR2 to explore the possible cosmological interpretations, including the first-order phase transition, domain walls and cosmic strings, for the recent discovered stochastic signal in PTA data sets. By computing the Bayes factors between these cosmological models and the fiducial SMBHB model (summarized in \Table{BFs}), we find that the first-order phase transitions and cosmic strings share comparable support as the SMBHB model, thus their possibilities as the the source of the signal in the PTA data sets cannot be excluded. 
However, the Bayes factor of the domain wall model versus the SMBHB model is $0.009$, indicating that domain walls are strongly disfavored by the combined data sets. This is a new implication different from what NANOGrav collaboration obtained when analyzing the new physics based on their own data set. A recent work also shows that domain wall interpretation is hardly compatible with the stochastic signal because it leads to the overproduction of primordial black holes \citep{Gouttenoire:2023ftk}. The  exploration on the parameter space of these models also provides us some interesting physical implications. For instance, we find that 1) when simultaneously considering both the sound waves and bubble collisions contributions to the GW spectrum in the first-order phase transition, it turns out that the bubble collisions contribute more dominantly than the sound waves. Additionally, this strong phase transition should take place at temperature below the electroweak phase transition of the Standard Model;
2) cosmic strings are more likely to have a low reconnection probability, with a $95\%$ upper limit of $p<6.68\times10^{-2}$ and a $99.9\%$ upper limit of $p<0.311$. It implies that the detected signal can only be explained by fundamental strings at a confidence level of $2\sigma$, but Dirichlet strings may also be a possible explanation within $3\sigma$ confidence level.

\begin{table}[!tbp]
    \centering
    \caption{\label{BFs}BFs of the power-law (PL), first order phase transition (PT), domain wall (DW), and cosmic string (CS) models compared to the SMBHBs model.}
    \begin{tabular}{c c c c c}
        \hline
        \hline
        Model & \quad PL &\quad PT &\quad DW &\quad CS \,\\
        \hline
        \quad BF & \quad $0.569$ & \quad\quad $0.799$ &\quad\quad $0.009$ &\quad\quad $1.699$\quad\, \\
        \hline
    \end{tabular}
\end{table}

\textbf{Note added.} While finalizing the manuscript, we notice that two parallel independent works \citep{Bian:2023dnv,Figueroa:2023zhu} also investigate the possible cosmological origin of the signal detected by PTAs, by comparing the Bayes factors between models. Our analyses differ from theirs in the sense that \citep{Bian:2023dnv} uses the NANOGrav, PPTA and EPTA data sets separately and employs the data from only the first five frequencies, \citep{Figueroa:2023zhu} uses the combined NANOGrav+EPTA data and considers models different from us. 

\textit{Acknowledgements.}
We are grateful to Lang Liu and Yang Jiang for helpful discussions.
We acknowledge the use of HPC Cluster of ITP-CAS. QGH is supported by the  grants from NSFC (Grant No.~12250010, 11975019, 11991052, 12047503), Key Research Program of Frontier Sciences, CAS, Grant No.~ZDBS-LY-7009, CAS Project for Young Scientists in Basic Research YSBR-006, the Key Research Program of the Chinese Academy of Sciences (Grant No.~XDPB15). 
ZCC is supported by the National Natural Science Foundation of China (Grant No.~12247176 and No.~12247112) and the China Postdoctoral Science Foundation Fellowship No.~2022M710429.

\bibliography{GW_sources}
\end{document}